\documentstyle[11pt,paspconf]{article}

\begin{document}

\title{Redshift-Space Density versus Real-Space Velocity Comparison} 
\author{Micha{\l} J.\ Chodorowski}
\affil{Copernicus Astronomical Center, Bartycka 18, 00--716 Warsaw, Poland}

\begin{abstract}
I propose to compare the {\em redshift-space\/} density field directly
to the {\em real-space\/} velocity field. Such a comparison possesses
all of the advantages of the conventional redshift-space analyses,
while at the same time it is free of their disadvantages. In
particular, the model-dependent reconstruction of the density field in
real space is unnecessary, and so is the reconstruction of the
velocity field in redshift space. The redshift-space velocity field
can be reconstructed only at the linear order, because only at this
order it is irrotational. Unlike the conventional redshift-space
density--velocity comparisons, the comparison proposed here does not
have to be restricted to the linear regime. Nonlinear effects can then
be used to break the $\Omega$--bias degeneracy plaguing the analyses
based on the linear theory. I present a degeneracy-breaking method for
the case of nonlinear but local bias.

\end{abstract}


\section{Introduction}
\label{sec:intro}

Comparisons between density fields of galaxies and the fields of their
peculiar velocities, a powerful tool to measure the cosmological
parameter $\Omega$, are commonly performed in real space. A necessary
ingredient of real-space analyses is the reconstruction of the galaxy
density field in real space from the observed redshift-space galaxy
field. This reconstruction is model-dependent: performing it, one has
to correct the redshifts for peculiar velocities of galaxies and the
amplitude of these velocities depends on $\Omega$. (Or, in the case of
inclusion of the bias between the galaxy and the mass distributions,
on $\beta$.)

To avoid this problem, the comparisons in redshift space have been
proposed (Nusser \& Davis 1994). However, the velocity field in
redshift space is irrotational only at the linear
order,\footnote{Chodorowski \& Nusser (1999) have shown that the
nonlinear redshift-space velocity field is irrotational in the distant
observer limit. However, the catalogs of peculiar velocities are not
yet deep enough to satisfy this assumption.} so that the
redshift-space analyses must be restricted to the linear regime. This
is unsatisfactory, since the derived amplitude of the density
fluctuations from current redshift surveys (e.g., Fisher {et
al.}~1995) and from the {\sc potent} reconstruction of density fields
(e.g., Dekel {et al.}~1990) slightly exceeds the linear regime.
Moreover, nonlinear effects may lead to breaking the degeneracy
between $\Omega$ and bias (Dekel {et al.}~1993, Bernardeau {et
al.}~1999).

Here I propose to compare the {\em redshift-space\/} density field
directly to the {\em real-space\/} velocity field. Such a comparison
enables one to avoid the model-dependent reconstruction of the density
field in real space. Also, the vorticity of the velocity field is no
longer a problem, because (before shell crossings) the real-space
velocity field is irrotational without any restrictions. Therefore,
the comparison proposed here combines all advantages of real-space and
redshift-space analyses, while at the same time it is free from their
disadvantages.

\section{Exact relation}
\label{sec:rel_ex}

The transformation from the real space coordinate, ${\mathbf r}$, to the
redshift space coordinate, ${\mathbf s}$, is (Kaiser 1987)
\begin{equation} 
{\mathbf s} = {\mathbf r} + v_r \hat{{\mathbf r}} \,,
\label{eq:rtos}
\end{equation}
where $\hat{{\mathbf r}} = {\mathbf r}/r$, $v_r({\mathbf r}) =
{\mathbf v} \cdot \hat{{\mathbf r}}$, ${\mathbf v}({\mathbf r})$ is
the real-space velocity field, and velocities are measured relative to
the Local Group. The relation between the redshift-space {\em
galaxy\/} density field, $\delta^{\rm (g)}_s$, and the real-space
velocity field is (Chodorowski 1999b)
\begin{equation}
\delta_{s}^{\rm (g)}({\mathbf s}) = \sum_{n = 0}^{\infty}
\frac{(-1)^n}{n!}  \left. \left\{ \frac{1}{r^2}
\frac{\partial^n}{\partial r^n} \left[ r^2 v_r^n \left(
\frac{\phi(r)}{\phi[s(r)]} \left[{\cal B} \circ {\cal F}(\partial
v_i/\partial r_j) + 1\right] - J \right) \right] \right\}
\right|_{\tilde{{\mathbf r}} = {\mathbf s}} .
\label{eq:dv_bias}
\end{equation} 
Here, $\phi$ is the selection function and $J$ is the Jacobian of the
transformation~(\ref{eq:rtos}), $J({\mathbf r}) = (1 + v_r/r)^2 (1 +
v_r')$, where $' \equiv \partial/\partial r$. The function $\cal F$
(e.g., Bernardeau 1999 and references therein) relates the real-space
velocity field to the real-space {\em mass\/} density field,
\begin{equation}
\delta({\mathbf r}) = {\cal F}[\partial v_i/\partial r_j({\mathbf
r}),\Omega] \,.
\label{eq:dvr_real}
\end{equation}
The function $\cal B$ relates locally the real-space mass density
field to the real-space galaxy density field, $\delta^{\rm
(g)}$, (the, so-called, local bias model),
\begin{equation}
\delta^{\rm (g)}({\mathbf r}) = {\cal B}\left[\delta({\mathbf r})\right]
\,. 
\label{eq:bias}
\end{equation}
If the selection function is given by a power law, $\phi \propto
s^{-p}$, then $\phi(r)/\phi[s(r)] = (1 + v_r/r)^p$. If not, the
expression $\phi(r)/\phi(s)$ should be expanded explicitly, and in any
case it will be a function of $v_r/r$. Therefore, both
$\phi(r)/\phi(s)$ and $J$ are functions of the real-space velocity
field and its derivatives, and equation~(\ref{eq:dv_bias}) is indeed a
relation between the redshift-space galaxy density and the real-space
velocity.

The key point of equation~(\ref{eq:dv_bias}) is that it relates the
redshift-space density at a point ${\mathbf s}$ to an expression
involving the real-space velocity field evaluated at $\tilde{{\mathbf
r}} = {\mathbf s}$. ($\tilde{{\mathbf r}}$ is a real-space point, in
general different from ${\mathbf r}$, which is related to ${\mathbf
s}$ by eq.~\ref{eq:rtos}.) This is very convenient, since now we can
treat the two fields as if they were given in the same coordinate
system.

\section{Approximate relation}
\label{sec:rel_app}
Relation~(\ref{eq:dv_bias}) involves an infinite series in velocity. If
the density--velocity comparison is performed on scales large enough
so that they are only weakly nonlinear, then this series can be
truncated.

{\bf Linear relation.} The linear density--velocity relation is

\begin{equation}
\delta_s^{\rm (g)}({\mathbf s}) = \left. \left[ - \beta^{-1}
\nabla_{{\mathbf r}} \cdot {\mathbf v} - v_r' - \left(2 + \frac{d \ln
\phi}{d \ln r} \right) \frac{v_r}{r} \right] \right|_{\tilde{{\mathbf
r}} = {\mathbf s}} \,,
\label{eq:de_s_1}
\end{equation}
where $\beta \equiv \Omega^{0.6}/b$ and $b$ is the linear bias
parameter. This equation coincides with equation~(6) of Nusser \&
Davis (1994).

{\bf Second-order relation.} To second order, the function $\cal F$ is
(e.g., Chodorowski 1997)

\begin{equation}
{\cal F} = - \Omega^{-0.6} \theta({\mathbf r}) + {\textstyle
\frac{4}{21}} \Omega^{-1.2} \left[\theta^2({\mathbf r}) - {\textstyle
\frac{3}{2}} \Sigma^2({\mathbf r}) \right] \,.
\label{eq:del_2}
\end{equation}
Here, $\Sigma^2 \equiv \Sigma_{ij} \Sigma_{ij}$, $\Sigma_{ij} \equiv
{\textstyle \frac{1}{2}} \left(\partial v_i/\partial r_j + \partial
v_j/\partial r_i \right) - {\textstyle \frac{1}{3}} \delta^{K}_{ij}
\theta$, and $\theta \equiv \nabla_{{\mathbf r}} \cdot {\mathbf
v}$. The symbol $\delta^{K}_{ij}$ denotes the Kronecker delta. The
function $\cal B$ is

\begin{equation}
{\cal B} = b \delta({\mathbf r}) + {\textstyle \frac{1}{2}} b_2
\left[\delta^2({\mathbf r}) - \langle \delta^2 \rangle\right] \,.
\label{eq:bias_2}
\end{equation} 
Here, $b_2$ is the nonlinear (second-order) bias parameter. This
yields
\begin{eqnarray}
\delta_s^{\rm (g)}({\mathbf s}) 
&\!\!\!=\!\!\!& 
\left\{
- \beta^{-1} \theta - v_r' - (2 + D_1) v_r/r 
+ \left[v_r(\beta^{-1} \theta + v_r')\right]'
\right.
\nonumber \\
&\!\!\!~\!\!\!&
+ \left(\frac{4}{21 b} + \frac{b_2}{2 b^2}\right) \beta^{-2}
\left(\theta^2 - {\textstyle \frac{3}{2}} \Sigma^2\right) 
+ (2 + D_1) \left(\beta^{-1} \theta + 2 v_r'\right) v_r/r 
\nonumber \\
&\!\!\!~\!\!\!&
\left. \left.
+ \left(1 + D_1 + D_1^2 - D_2 \right)^{\vphantom{2}} (v_r/r)^2
\right\} \right|_{\tilde{{\mathbf r}} = {\mathbf s}} \,, 
\label{eq:dv_bias_2}
\end{eqnarray}
where $D_1 = d \ln \phi/d \ln r$, and $D_2 = (\phi'' r^2)/(2 \phi)$. 

Equation~(\ref{eq:dv_bias_2}) can be used to reconstruct the
real-space velocity field from the associated redshift-space galaxy
density field. Since the real-space velocity field is irrotational, it
can be described as a gradient of the velocity potential,

\begin{equation}
{\mathbf v}({\mathbf r}) = - \nabla_{{\mathbf r}} \Phi_v \,.
\label{eq:vel_pot}
\end{equation}
Equation~(\ref{eq:dv_bias_2}) reduces then to a nonlinear differential
equation for the velocity potential, which can be solved
iteratively. First, we solve its linear part. Next, we find the
second-order solution by solving again the linear equation, with the
source term resulting from the density modified by nonlinear
contributions approximated by first-order solutions. Specifically,
\begin{equation}
\beta^{-1} \Delta_{{\mathbf r}} \Phi_v^{(2)} + \frac{\partial^2
\Phi_v^{(2)}}{\partial r^2} + \frac{2 + D_1}{r} \frac{\partial
\Phi_v^{(2)}}{\partial r} = \delta_s^{\rm (g)}(\tilde{{\mathbf s}} =
{\mathbf r}) - {\cal N}_2\left[\Phi_v^{(1)}({\mathbf r})\right] \,,
\label{eq:pot_bias}
\end{equation}
where ${\cal N}_2$ is a sum of all terms quadratic in velocity in
equation~(\ref{eq:dv_bias_2}), expressed as functions of the potential. 

\section{Breaking the ${\mathbf \lowercase{\Omega}}$--bias degeneracy}
\label{sec:Omega_bias}

Based on~(\ref{eq:pot_bias}), from the associated redshift galaxy
field we can reconstruct the nonlinear real-space velocity field. The
latter can then be compared to measured radial velocities of
galaxies. This comparison will yield the best-fit values of two
parameters: $\beta$, and
\begin{equation}
\beta_2 \equiv \left(\frac{4}{21 b} + \frac{b_2}{2 b^2}\right)^{-1} 
\beta^2 
\label{eq:beta_2}
\end{equation}
(see eq.~\ref{eq:dv_bias_2}). They are a combination of three physical
parameters: $\Omega$, $b$, and $b_2$. Therefore, we need an additional
constraint on these parameters. As this constraint one can adopt the
large-scale galaxy density skewness (Bernardeau {et al.}~1999).

We can measure the redshift {\em galaxy\/} density skewness, $S_{3
s}^{\rm (g)}$, (e.g., Kim \& Strauss 1998), while gravitational
instability theory can predict the value of the redshift {\em mass\/}
density skewness, $S_{3 s}$, (Hivon {et al.}~1995). The relation
between the two is approximately (for details see Chodorowski 1999a)
\begin{equation}
S_{3 s}^{\rm (g)} = \frac{S_{3 s}}{b} + 3 \frac{b_2}{b^2} 
\,.
\label{eq:S_3}
\end{equation}
Now we have three independent constraints for the parameters $\Omega$,
$b$ and $b_2$, so that in principle we can measure $\Omega$ and bias
separately. 

The additional constraint on $\Omega$ and bias (the skewness) is to be
inferred from the density field alone, making any additional
observations unnecessary. In conclusion, there is enough information
in the density field and the associated velocity field to break the
$\Omega$--bias degeneracy.

For a more detailed discussion of the problem, see Chodorowski
(1999b).

\acknowledgments
This research has been supported in part by the Polish State Committee
for Sci.\ Research grants No.~2.P03D.008.13 and 2.P03D.004.13.

\end{document}